\documentclass[twoside,twocolumn,showpacs,floatfix]{revtex4}
\usepackage{amssymb,amsmath}
\usepackage[pdftex]{graphicx}

\begin{document}

\title{On the dynamics of traveling phase-oscillators with positive and negative couplings}

\author{Jungzae \surname{Choi}}
\affiliation{Department of Physics and Department of Chemical Engineering, Keimyung University, Daegu 704-701, Korea}
\author{MooYoung \surname{Choi}}
\affiliation{Department of Physics and Astronomy and Center for Theoretical Physics, Seoul National University, Seoul 151-747, Korea}
\author{Byung-Gook \surname{Yoon}}
\thanks{Author to whom correspondence should be addressed. E-mail: bgyoon@ulsan.ac.kr}
\affiliation{Department of Physics, University of Ulsan, Ulsan 680-749, Korea}

\begin{abstract}

We investigate numerically the dynamics of traveling clusters in systems of
phase oscillators, some of which possess positive couplings and others negative couplings. The phase distribution, speed of traveling, and average separation between clusters as well as order parameters for positive and negative oscillators are computed, as the ratio of the two coupling constants and/or the fraction of positive oscillators are varied. The traveling speed depending on these parameters is obtained and observed to fit well with the numerical data of the systems. With the help of this, we describe the conditions for the traveling state to appear in the systems with or without periodic driving.

\end{abstract}

\maketitle

\section{Introduction}

The Kuramoto model provides a convenient starting point for studying collective synchronization in phase oscillators~\cite{ref:Kuramoto0,ref:Kuramoto}. After earlier studies, there have appeared many extensions and variations of the Kuramoto model~\cite{ref:extension}. Some of them introduce repulsive or negative couplings to all~\cite{ref:repulsive} or fractions of oscillators \cite{ref:hs1,ref:hs2, ref:lai}.

When there are repulsive as well as attractive couplings between oscillators and each
oscillator is identified to be either repulsive or attractive, there may appear two traveling clusters separated by a phase difference less than $\pi$ radian~\cite{ref:hs1}. %However, no further work has been done
%on the dynamics of the traveling state.
This state arises in the system with weak disorder, when the magnitude of the repulsive or negative coupling is smaller than that of the positive coupling and the numbers of the two type of oscillators are not too different.
However, no further work has been reported on the dynamics of the traveling state.

This work considers the same oscillator system with positive and negative couplings, as considered in Ref.~\cite{ref:hs1} and investigates emergence of the traveling state. In particular, we obtain the traveling speed of clusters depending on relevant parameters of the system, which allows to describe the conditions for the emergence of the travelings state. %We further take into consideration a few systems with periodic driving, and
%the dynamics of the traveling state in these systems are discussed in terms of those conditions.

This paper consists of four sections: In Sec. II, the oscillator model and its dynamics are described. Section III presents numerical results together with the phenomenological interpretation of traveling and non-traveling states. Finally, a brief summary is given in Sec. IV.

\section{Model and Numerical calculation}

We consider a system of $N$ oscillators, the $i$th of which has intrinsic frequency $\omega_i$. It is described by
its phase and coupled globally to other oscillators.
The dynamics of such a coupled oscillator system is governed
by the set of Langevin equations of motion for the phase $\phi_i$
of the $i$th oscillator ($i=1,...,N$):
\begin{equation} \label{model}
 \dot{\phi_i} =\omega_i -  \frac{K_i}{N}\sum_{j=1}^N  \sin(\phi_i-\phi_j)+ I_{i} (t),
\end{equation}
where the intrinsic frequencies are assumed to be symmetrically distributed according the
Lorentzian distribution
$g(\omega)= (\gamma_{\omega}/\pi) (\omega^2 + \gamma_{\omega}^2 )^{-1}$.
The second term on the right-hand side represents
sinusoidal interactions with other oscillators, where the coupling constant $K_i$
takes a positive or negative value depending on
whether or not the oscillator $i$ follows the mean field [$\Delta$ in Eq.~(\ref{eqn}) below].
Specifically, the coupling is taken from the distribution
$\Gamma(K)=p\delta(K{-}K_{+})+(1-p)\delta(K{-}K_{-})$, where
$K_{\pm}$ is the positive/negative coupling constant ($K_{+} >0$ and $K_- <0$)
and $p$ is the fraction of oscillators having the positive coupling constant.
The last term describes the periodic driving.

In order to measure the synchronization of the system, we
introduce the complex order parameter:
\begin{equation} \label{deforder}
  \Psi \equiv \frac{1}{N} \sum_{j=1}^N e^{i \phi_j}
       = \Delta e^{i\theta} ,
\end{equation}
which characterizes synchronization of the oscillators, with the magnitude $\Delta$ and the average phase $\theta$. The order parameter defined in
Eq.~(\ref{deforder}) allows us to reduce Eq.~(\ref{model}) to a
{\em single} decoupled equation:
\begin{equation} \label{eqn}
 \dot{\phi_i} = \omega_i - K_i \Delta \sin(\phi_i -\theta) + I_{i} (t).
\end{equation}

To investigate the behavior of the system governed by Eq. (\ref{eqn}), we resort mainly to numerical methods.
Using the second-order Runge-Kutta-Helfand-Greenside algorithm, we integrate
Eq.~(\ref{eqn}) with the time step $\Delta t=0.01$ for the system size
$N=2000$. Initially ($t=0$), $\phi_i$'s are %either set to be zero or
randomly distributed between $0$ and $2\pi$ for all $i$. We fix the period
of driving, if present, to be $\tau=5.12$ and the positive coupling constant, $K_{+} =1$.

After the initial transient behavior,
the system reaches its stationarity, and we obtain the time series of the
order parameter information on the time evolution of the phase distribution.
In order to understand the clustering behavior,
we define the order parameter of positive/negative oscillators $\Psi_{\pm} = \Delta_{\pm} e^{i\theta_{\pm}}$ similarly to Eq.~(\ref{deforder}). These parameters as well as the phase split $\delta$,
which stands for the angular distance between the two phases $\theta_+$ and $\theta_-$ for positive and negative oscillators, respectively ($\delta \equiv |\theta_+ -\theta_- |$), are also calculated in each run.

The average values of these as well as the traveling speed $w$ are obtained as follows:
At each time, we calculate these parameters and the average value of phase velocities $w_i=\dot{\phi}_i$ over oscillators.
Then we get the time average over 10 periods of driving and take the absolute value to obtain the traveling speed.
When there is no driving, the order parameters $\Delta$'s do not vary much in time, and we sometimes take the values after the time evolution.
Finally, we take the averages over 30 initial configurations to obtain the (average) speed $w$.  % $\langle w \rangle$.

We also examine populations in given regions of the phase space.
Specifically, we divide one cycle of the phase angle into
72 different ranges, the $k$th of which is defined by the phase
interval $(k\pi/36{-}\pi/72, \,k\pi/36{+}\pi/72]$ (modulo $2\pi$), and obtain
the number $n_k$ of oscillators belonging to the $k$th range
($k =0,...,71$).

\section{Results and discussion}

\begin{figure}
\includegraphics[width=8.5cm]{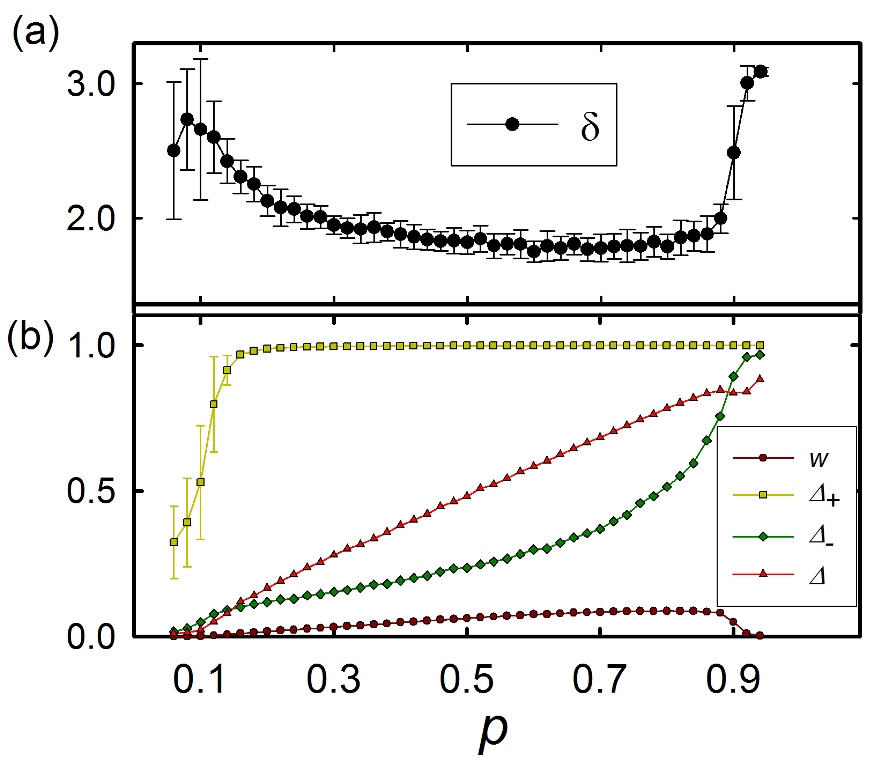}
%\caption{(color online) Average traveling speed $w$, %$\langle w \rangle$,
%phase split $\delta$, %$\langle \delta \rangle$ between average phase of positive oscillators, $\theta_+$, and that of negative ones %$\theta_-$,
%order parameters $\Delta_+$, % $\langle \Delta_+  \rangle$ of positive oscillators and
%$\Delta_-$, and %$\langle\Delta_- \rangle$ of negative ones, order parameter
%$\Delta$ %$\langle\Delta \rangle$ of the system
%versus the positive oscillator fraction $p$ in a system of $N=2000$ oscillators
\caption{(color online) (a) Phase split $\delta$ and (b) average traveling speed $w$, order parameters $\Delta_+$, $\Delta_-$, and $\Delta$  versus the fraction of positive oscillators $p$ in a system of $N=2000$ oscillators
with $K_- =-0.1$ and $\gamma_{\omega}=0.01$.
Error bars represent standard deviations and lines are merely guides to the eye.
}
\label{fig:wDp}
\end{figure}

When the coupling constant of one oscillator becomes negative in an ordered system of positive oscillators,
it should be repelled by other oscillators. After the initial transient, the phase split $\delta$ will become $\pi$ radian.
When clusters of positive and negative oscillators according to the delta distribution are introduced in the phase space and if $|K_-|$ is less than $K_+$, the phase split $\delta$ usually
decreases at first, increases subsequently, and so on, finally reaching a quasi-stationary state in which it fluctuates around a constant value.
Especially, whenever $|K_-|$ is larger than $K_+$, the phase gap $\delta$ rapidly increases to $\pi$; this is called the $\pi$-state in Ref.~\cite{ref:hs1}.
To explain how this behavior depends on the parameters of the system requires understanding the dynamics of clusters before the quasi-stationary state is formed.

%%%following sentece modified
Figure~\ref{fig:wDp}(a) and (b) displays the phase split $\delta$ between the average phases $\theta_{\pm}$ of positive/negative oscillator and the average values of the traveling speed $w$ %$\langle w \rangle$,
as well as the order parameters $\Delta_{\pm}$ and $\Delta$ of positive/negative oscillators and of the whole system, respectively,
versus the fraction $p$ of positive oscillators
in the system of $N=2000$ oscillators with $K_- =-0.1$ and $\gamma_{\omega}=0.01$.
As $p$ is increased, the order parameter $\Delta_+$ grows rapidly to values close to unity ($\Delta_+ \approx 1$) while $\Delta_-$ grows slowly. Meanwhile, the phase split decreases at first, to display rather a wide valley, and then rises rapidly.
%Here it is of interest to note that the presence of a small fraction of negative oscillators of weak strength induces positive oscillators to be %ordered more. As the %raction $p$ is increased, the phase split reduces, reaching the minimum value.
%Here it is of interest to note that the order parameter of positive oscillators increase rapidly starting from very small value
%of $p$ in the presence of majority opposite ocillators of weak strength.
%As the fraction $p$ is increased, the phase split reduces, reaching the minimum value.
%To meet the refree's checkpoint
%Note again that he overall order parameter decreases a little at the region near $p \approx 0.9$, which is caused
%by the rapid increase in the phase split. As the two clusters are separated rapidly, the order parameter will be decreased, and this effect is dominat over that of %increasing $p$.
Here it is of interest to note that the order parameter of positive oscillators increases rapidly as $p$ is increased from a small value in the presence of a majority of opposite oscillators; this is possible because the coupling ratio $|K_{-}|/K_+$ is very small. 
Note also that the overall order parameter exhibits a slight decrease around $p \approx 0.9$, which is caused by the rapid increase in the phase split. As the two clusters are separated rapidly, the order parameter should decrease 
and this effect is dominant in the narrow region around $p\approx 0.9$. 
For still larger values of $p$, the increasing effect of positive oscillators becomes dominant
and $\Delta$ increases again.

\begin{figure}
\includegraphics[width=8.5cm]{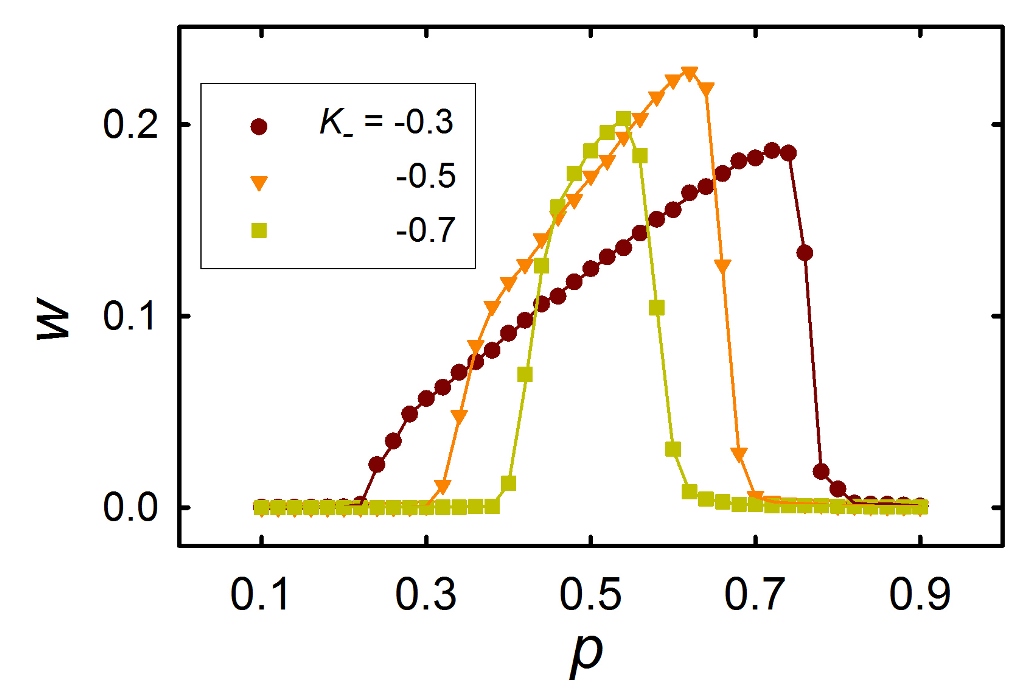}
\caption{(color online) Average traveling speed $w$ versus the fraction $p$ of positive oscillators in a system of $N=2000$ oscillators
with $\gamma_{\omega}=0.01$ and $K_- = -0.3, -0.5$ and $-0.7$ as shown in the legend.
Lines are plots of Eq.~(\ref{wformula}) whereas symbols plot the data points obtained numerically. Excellent agreement is observed.
}
\label{fig:wp}
\end{figure}

The traveling speed begins to increase
from zero, together with the first rise of $\Delta_+$ and the rapid decrease of $\delta$.
It keeps growing until the sudden decay with the rapid increase of the phase split.
Note that $\Delta_+$ increases rapidly with $p$ in this region. Similar behaviors
of $w$ are observed in systems with larger values of  $|K_{-}|$, as shown in Fig.~\ref{fig:wp}, although the rise/decay starts at larger/smaller values of $p$. Here
only the data on the traveling speed are presented; other quantities
behave similarly to the corresponding data in Fig.~\ref{fig:wDp}.

Such behaviors can be explained as follows:
Suppose that there are $Np \Delta_+$ positive oscillators having the phase $\theta_+$
and $N(1-p) \Delta_-$ negative oscillators with the phase $\theta_-$ and assume that
the remaining oscillators are desynchronized and have no net effect on $w$.
With this assumption, it is straightforward to derive, from Eq.~(\ref{model}),
\begin{equation} \label{wformula}
     w  =  \Delta_+ \Delta_- (|K_-|+K_+ ) p(1-p) \sin \delta ,
\end{equation}
which gives a good description of the data on the traveling speed.
For example, the lines in Fig.~\ref{fig:wp}, plotting Eq.~(\ref{wformula}),
describe excellently the data points represented by symbols.
Unless distributions of the two clusters are very broad or the order parameters $\Delta_+$
and $\Delta_-$ are very small, Eq.~(\ref{wformula}) yields good fitting with data points.
%Of course, the values of order parameters and the phase separation must be supplied through numerical integration, in order to use this formula. %%%
This equation manifests that the initial increase of $w$ is due to the increase of
$\Delta_-$  as well as to the decrease of $\delta$. The final decrease is attributed to the rapid increase of the phase split to $\pi$.

Equation~(\ref{wformula}) indicates that the traveling speeds $w_{\pm}$ of positive and negative oscillators can be written in the form:
\begin{eqnarray} %\label{wformula+}
     w_+ &=& \Delta_+ \Delta_-  K_+  (1-p) \sin \delta \nonumber \\
     w_- &=& \Delta_+ \Delta_- |K_{-}| p \sin \delta.
\label{wformula-}
\end{eqnarray}
It should be pointed out that in the near-stationary state, the traveling speed of positive oscillators, when clustered, is almost the same as the phase speed $\dot{\theta}_{+}$ of the average phase of positive oscillators while that of negative oscillators is smaller
than $\dot{\theta}_-$. The reason why the traveling speed of a negative cluster
can be lower than its phase speed $\dot{\theta}_-$ may be understood as follows:
Negative oscillators repel each other in a cluster and the negative cluster, if left alone, becomes unstable.
Due to this unstableness and attraction of positive oscillators,
some negative oscillators tend to move away from the peak of the cluster
toward the positive cluster.
This will make the traveling speed less than the phase speed of the negative cluster.

Now, using Eqs.~(\ref{wformula}) and (\ref{wformula-}), we can probe how the $\pi$-state forms from the two clusters of the delta distribution, namely, positive oscillators at one given phase and negative oscillators at another given phase.
If these clusters are separated initially by a phase difference less than $\pi$,
the negative cluster will run away from the positive one, with the latter following the former.
When the phase speed of negative oscillators is larger than that of positive ones, the phase split increases to $\pi$ radian.
Then all traveling speeds vanish together and a non-traveling
stationary state is formed. %The reason why the traveling speed of a negative cluster
%can be lower than its phase speed $\dot{\theta}_-$ may be understood as follows:
%Negative oscillators repel each other in a cluster and the negative cluster, if left alone, becomes unstable.
%Due to this unstableness and attraction of positive oscillators,
%negative oscillators tend to move away from the peak of the cluster
%toward the positive cluster.
%This will make the traveling speed less than the phase speed of the negative cluster.

The traveling state can be understood in a similar manner, again in consideration of the two clusters of the delta distribution.
%When the traveling speed of the positive cluster is larger than $\dot{\theta}_-$, the phase split may never reach $\pi$ radian and will arrive in a quasi-stationary state where the phase speeds of the two clusters are almost the same (and larger than the traveling speed of negative oscillators.)
When the traveling speed of the positive cluster is sufficiently larger than that of the negative cluster,
the phase split may never reach $\pi$ radian and will arrive in a quasi-stationary state where the phase speeds of the two clusters are almost the same (and larger than the traveling speed of negative oscillators.)
Here the positive cluster appears to chase the negative one, making it move faster than its traveling speed,
which has been confirmed (data not shown). However, as stated before, the actual transient behaviors may be quite complicated,
sometimes exhibiting oscillations of the phase split.

\begin{figure}
\includegraphics[width=8.5cm]{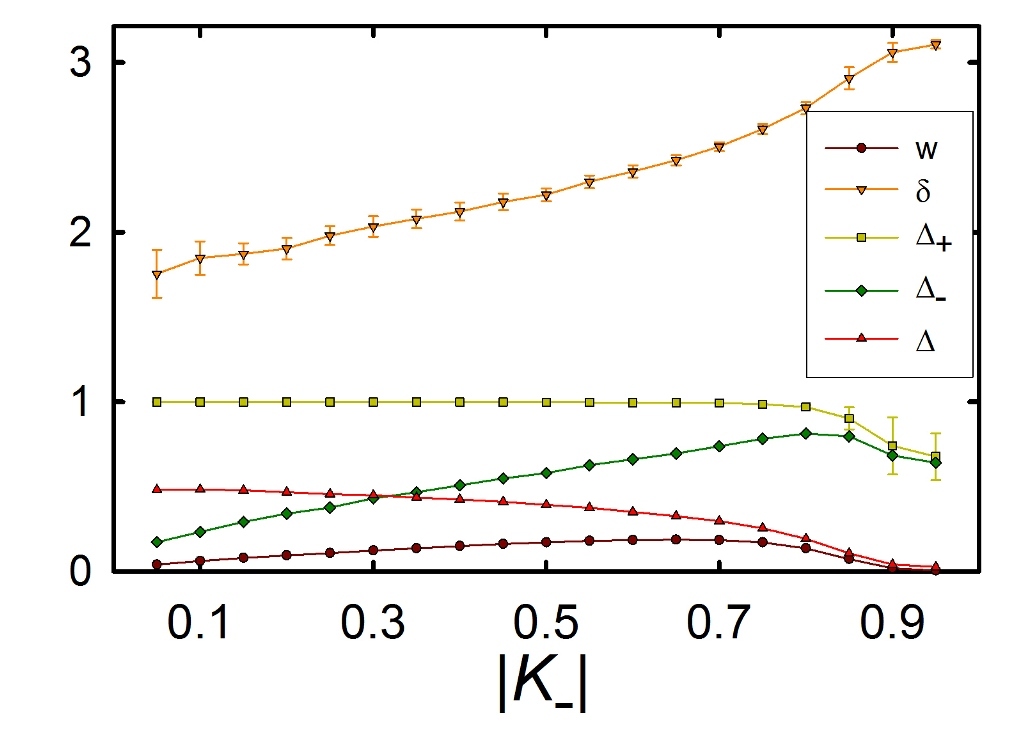}
\caption{(color online) Average traveling speed $w$, %$\langle w \rangle$,
phase split $\delta$, %angle $\langle \delta \rangle$,
order parameters $\Delta_+$, $\Delta_-$, and $\Delta$
%$\langle \Delta_+  \rangle$ of positive oscillators and
%$\langle\Delta_- \rangle$ of negative ones, order parameter $\langle\Delta \rangle$ of the system
versus the magnitude of the negative coupling constant $|K_-|$ in a system of $N=2000$ oscillators with $p =0.5$ and $\gamma_{\omega}=0.01$.
Error bars represent standard deviations and lines are merely guides to the eye.
}
\label{fig:wDK}
\end{figure}

\begin{figure}
\includegraphics[width=8cm]{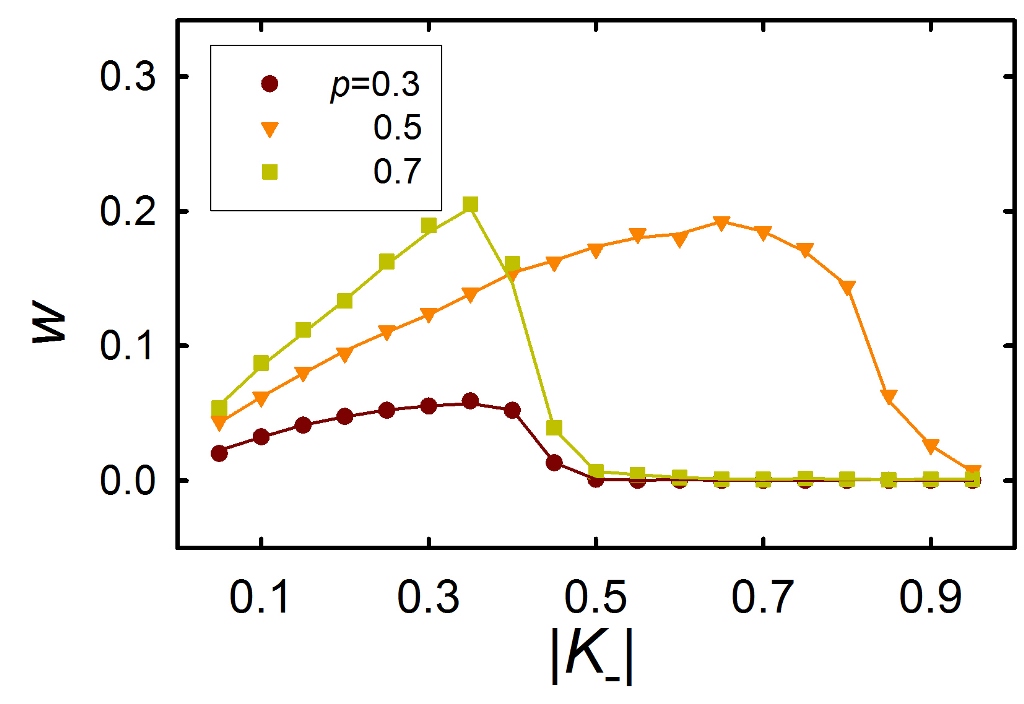}
\caption{(color online) Average traveling speed $w$ %$\langle w \rangle$
versus the magnitude of the negative coupling constant $|K_{-}|$
in a system of $N=2000$ oscillators with $\gamma_{\omega}=0.01$ and $p = 0.3, 0.5$, and $0.7$, as shown in the legend.
Lines are plots of Eq.~(\ref{wformula}) whereas symbols plot the data points obtained numerically. Again excellent agreement is observed.
}
\label{fig:wK}
\end{figure}

With this in mind, one can interpret the obtained data further. When $|K_-|p$ is larger than $ K_+ (1-p)$, there may not arise the traveling state since $\dot{\theta}_-$ is larger than  $\dot{\theta}_+$.
%This condition explains qualitatively the upper limits of the traveling state in Fig.~\ref{fig:wp}, which decline with $|K_-|$ increasing.
This condition is consistent with the condition for the $\pi$-state given by Eq. (15) in Ref.~\cite{ref:hs1}, and
explains qualitatively the upper limits of the traveling state in Fig.~\ref{fig:wp}, which decline with $|K_-|$ increasing.

We now examine the dependence upon the strength of negative coupling $|K_{-}|$.
We thus consider a system with $p$ set to be $0.5$, and compute the same (ensemble)
average quantities as in Fig.~\ref{fig:wDp}.
The obtained data versus $|K_{-}|$ are presented in Fig.~\ref{fig:wDK}.
When the numbers of negative and positive oscillators are nearly equal, the phase split increases with $|K_-|$ as expected. Here again, $w$ increases at first, and then decreases. As for the initial increase, effects of increasing $\Delta_-$ dominate although increasing $\delta$ plays a role against this. The subsequent decrease is accompanied by both the decrease in order parameters and the increase in $\delta$ toward $\pi$; this can be understood with the help of Eq.~(\ref{wformula}).
At smaller or larger values of $p$, the transition to a non-traveling state takes place at smaller values of $|K_- |$, as shown in Fig.~\ref{fig:wK}, where Eq.~(\ref{wformula}) is plotted by lines.
When $p=0.3$, the number of positive oscillators is far smaller than that of negative oscillators. As the negative coupling increases above a threshold, the order parameter $\Delta_+$ decreases rapidly, which causes the traveling state to be less probable.
At the larger value of $p=0.7$, the value of $(1-p)K_+ $ is small and the traveling speed  $w_-$ becomes larger than $w_+$ at smaller values of $|K_-|$.

%discussion using two delta clusters is moved just after the paragraph containing the w formula.
%%%%%%%%%%%%%%%%%%%%%%%%%%%%%%%%%%%%%%%%%%%%%%%%%%%%%%%
%Now, using  Eq.~(\ref{wformula}), we can state how $\pi$-state is formed starting from two clusters of
%delta distribution, which means positive oscillators are at one angle and the other ocsillators are at another angle.
%Equation~(\ref{wformula}) means that traveling speed of positive (negative) oscillators, $w_+$ ($w_-$) is given as following:
%.
%.
%has been confirmed, although the data is not shown here. However, as stated before, the
%actual transient behaviors may be very complicated, sometimes exhibiting the oscillation of $\delta$.

\begin{figure}
\includegraphics[width=8.5cm]{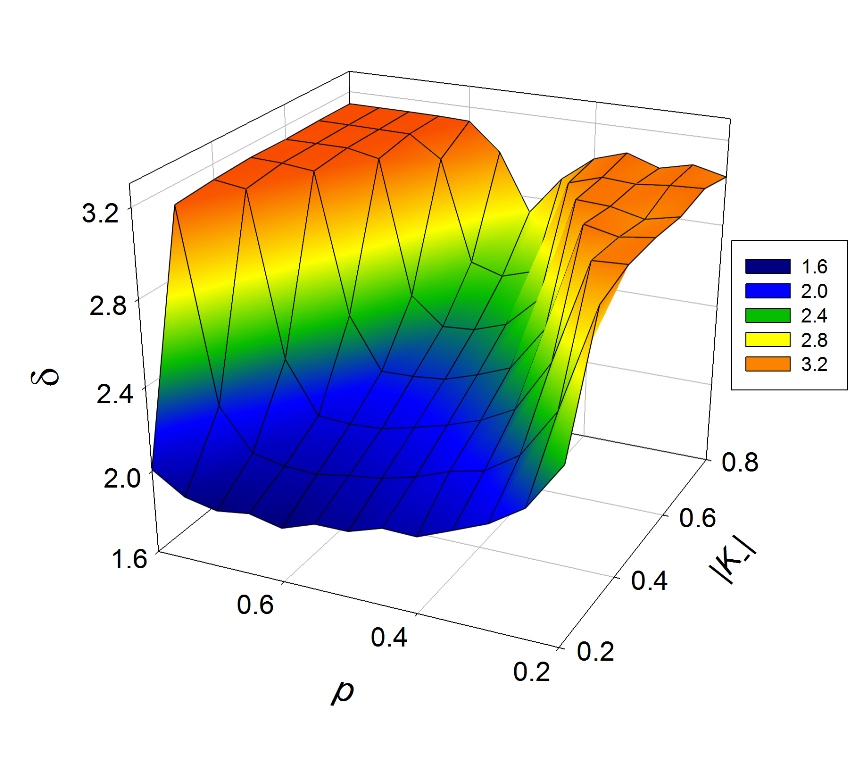}
%\caption{ Separation angle $\delta$ vs $(p, K_{-})$ for systems with $N=2000$ and $\gamma_{\omega}=0.01$.
\caption{Plot of the phase split $\delta$ on the plane $(p, |K_{-}|)$ for a system of $N=2000$ oscillators with $\gamma_{\omega}=0.01$.
}
\label{fig:dpK}
\end{figure}

Obviously, the most important condition for the traveling state is that two clusters should be formed and the phase split of clusters should be less than $\pi$ radian. Figure~\ref{fig:dpK} presents the three-dimensional plot of the phase split $\delta$ on the plane $(p, |K_{-}|)$ %$(p, K_{-})$
for a system of $N=2000$ oscillators with $\gamma_{\omega}=0.01$ in the absence of driving.
If the dispersion in natural frequencies is raised,
%the region of traveling state is shrinked. Similary, to much noise, if present, will make the clusters
the domain of the traveling state shrinks. Similarly, strong noise, if present, will make the clusters broadened or disappear, inhibiting the traveling state from forming.

\begin{figure}
\includegraphics[width=8.5cm]{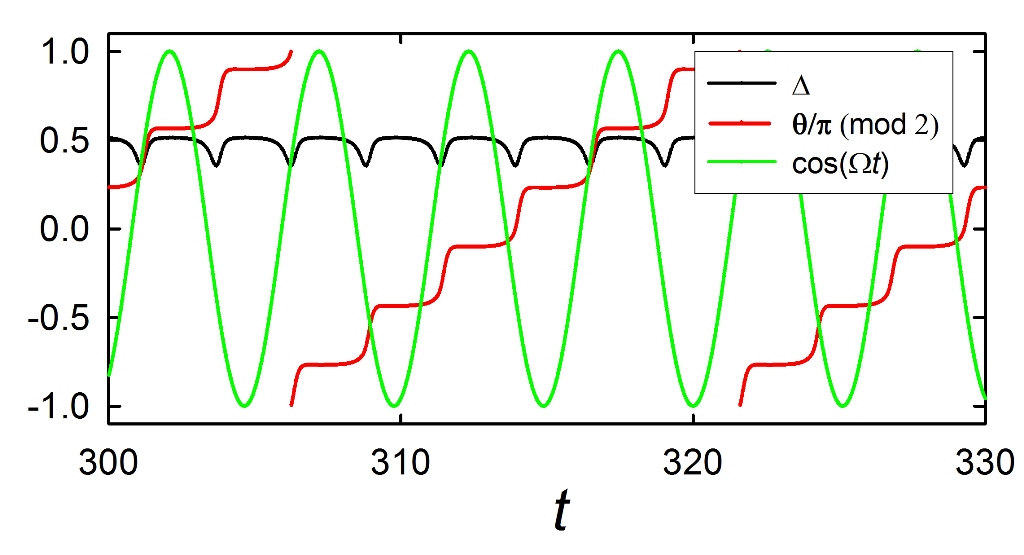}
\caption{(color online) Time evolution of the magnitude $\Delta$ and phase $\theta/\pi$ (modulo 2) of the order parameter for a system of $N=2000$ oscillators with $p=0.6$, $K_{-} =-2.0$ and $\gamma_{\omega}=0.01$. The system is
driven by the periodic three-fold symmetry breaking field $I_i (t)= h_0 \sin(3\phi_i ) \cos (2\pi t/\tau)$ with $h_{0}=5$ and $\tau=5.12$.
Also plotted is the cosine function of the driving.
}
\label{fig:DtH}
\end{figure}

\begin{figure}
\includegraphics[width=8.5cm]{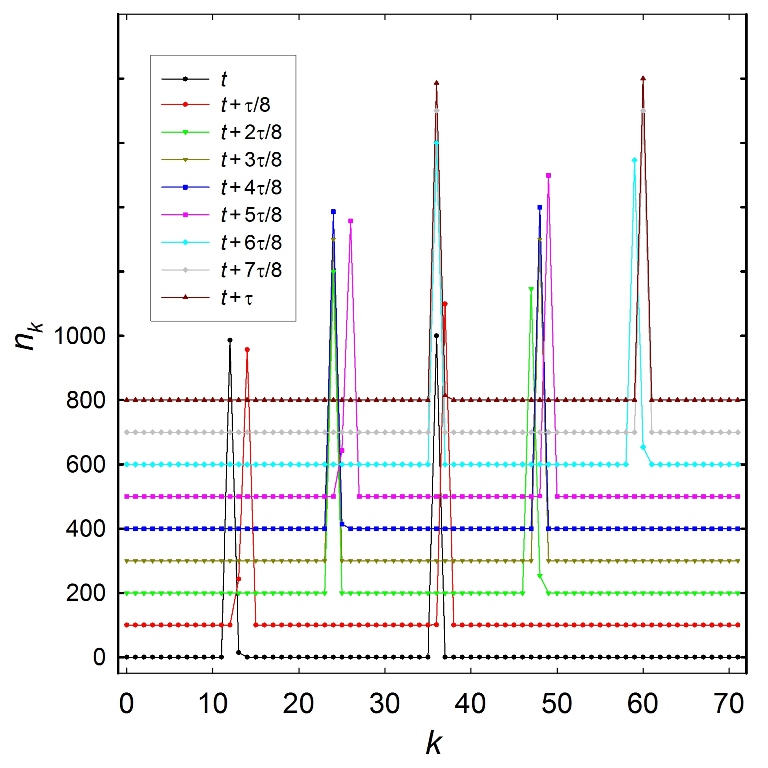}
\caption{(color online) Number $n_k$ of oscillators belonging to range $k$
at nine successive times as shown in the legend. Data have been obtained from the same
system as in Fig.~\ref{fig:DtH}. Here, an oscillator is considered
to be in range $k$ if its phase $\phi$ satisfies $k\pi/36{-}\pi/72 < \phi \leq k\pi/36{+}\pi/72$ (modulo $2\pi$). The lowermost curve describes $n_k$ versus $k$ at the earliest time $t$ while the uppermost one at the time after one period of driving from $t$. For clarity, the curves have been shifted upward by $100$ successively.
}
\label{fig:disth5}
\end{figure}

Another condition for the traveling state to emerge is simply the condition necessary for the phase split to be less than $\pi$.  For systems without driving, the magnitude of negative coupling
%mmust be smalller than $K_+$. However, the relative strength of two coupings is not an absolute one.
must be smaller than $K_+$ for traveling state to be formed. However, this condition for the relative strength of two coupling constants is not an absolute one.
To show this, we consider a system driven by a periodic three-fold symmetry breaking field: %$I_i (t)= h_0  \sin(3 \phi_i ) \cos ( 2 \pi t/\tau$ with $\tau=5.12$. Figure~\ref{fig:DtH} shows
$I_i (t)= h_0  \sin(3 \phi_i ) \cos ( 2 \pi t/\tau)$ with $\tau=5.12$. We choose the absolute value of $K_{-}\,(=2)$ to be larger than $K_{+}\,(=1)$,
for which the traveling state cannot emerge in the absence of driving.
Figure~\ref{fig:DtH} shows how the magnitude $\Delta$ and phase $\theta/\pi$ (modulo 2) of the order parameter evolve in time $t$ for a system of $N=2000$ oscillators with $p=0.6$, $\gamma_{\omega}=0.01$, and $h_0 =5$.
When the driving amplitude $h_0$ is sufficiently large, two clusters form,  %and two clusters are formed
%separated by $\approx 2\pi/3$ radian. As is seen in the figure, traveling does occur, and after three %%
separated by $2\pi/3$ radian. See Fig.~\ref{fig:disth5}, which shows the
number $n_k$ of oscillators belonging to range $k$ at nine successive times. The lowermost curve exhibits the number at the earliest time $t$; the uppermost one at the time after one period of driving from $t$.
As manifested in Figs.~\ref{fig:DtH} and \ref{fig:disth5}, there emerges traveling and after three periods of driving, the phase of the system increases by $2\pi$. % Note the absolute value value of  $K_{-}(=2)$ is larger than  $K_{+}(=1)$,
%and the traveling states cannot happen for this value of  $K_{-}$ in a system of no driving,

\section{Summary}

%We have studied, by means of extensive numerical calculations, an oscillator system with positive and negative couplings.
%The traveling speeds depending on appropriate parameters have been obtained and observed to fit well with the numerical data. With the help of this, we describe the dynamics of the traveling state in the system.
We have studied, by means of numerical calculations, an oscillator system with positive and negative couplings, and investigated emergence of the traveling state.
In particular, we have obtained traveling speeds of clusters depending on relevant parameters of the system, in terms of which the condition for the emergence of the traveling state has been expressed:
If the traveling speed of a negative cluster is sufficiently smaller than that of a positive cluster, the phase split of two clusters may never reach $\pi$ radian and the positive cluster permanently chases the negative one,
resulting in the traveling state.

\acknowledgments
This work was supported in part by the 2013 Research Fund of University of Ulsan.


\begin{thebibliography}{99}
%\bibitem{ref:Winfree}
%For a list of references, see A.T. Winfree, {\it The Geometry of Biological
%Time} (Springer-Verlag, New York, 1980); J. Theor. Biol. \textbf{16}, 15 (1967).

\bibitem{ref:Kuramoto0}
Y. Kuramoto, {\it Chemical Oscillations, Waves, and Turbulence}
(Springer-Verlag, Berlin, 1984).

\bibitem{ref:Kuramoto}
Y. Kuramoto, in {\it Proceedings of the International Symposium on
Mathematical Problems in Theoretical Physics}, edited by H.
Araki (Springer-Verlag, New York, 1975);
Y. Kuramoto and I. Nishikawa, J. Stat. Phys. \textbf{49}, 569 (1987).

\bibitem{ref:extension}
See, e.g., H. Sakaguchi, Prog. Theor. Phys. \textbf{79}, 39 (1986);
J. L. Rogers and L. T. Wille, Phys. Rev. E \textbf{54}, R2193 (1996);
M. K. S. Young and S. H. Strogatz, Phys. Rev. Lett. \textbf{82}, 648 (1999);
S.C. Young and L.S. Tsimring,  Phys. Rev. E \textbf{65}, 041906 (2002).

\bibitem{ref:repulsive}
See, e.g.,  L. S. Tsimring, N. F. Rulkov, M. L. Larsen, and M. Gabbay,
 Phys. Rev. Lett. \textbf{95} 014101 (2005), and references therein.

\bibitem{ref:hs1}
H. Hong and S. H. Strogatz, Phys. Rev. Lett. \textbf{106}, 054102 (2011).

\bibitem{ref:hs2}
H. Hong and S. H. Strogatz, Phys. Rev. E \textbf{85}, 056210 (2012).

\bibitem{ref:lai}
Y. M. Lai and M. A. Porter, Phys. Rev. E \textbf{88}, 012905 (2013).

%\bibitem{ref:psynch}
%M. Y. Choi, Y. W. Kim and D. C. Hong, Phys. Rev. E \textbf{49}, 3825 (1994);
%H. Hong, M.Y. Choi, B.-G. Yoon, K. Park, and K.-S. Soh, J. Phys. A \textbf{32}, L9 (1999);
%H. Hong and M.Y. Choi, Phys. Rev. E \textbf{62}, 6462 (2000).

% \bibitem{ref:inertia}
%H. Hong, M.Y. Choi, J. Yi, and K.-S. Soh, Phys. Rev. E \textbf{59}, 353 (1999);
%H. Hong and M.Y. Choi, Phys. Rev. E \textbf{62}, 6462 (2000).

%\bibitem{ref:cccy}
%J. Choi, M.Y. Choi, M.S. Chung and B.-G. Yoon, Int. J. Mod. Phys. B \textbf{27},
%1350062 (2013).

%\bibitem{ref:ccy14}
%J. Choi, M.Y. Choi and B.-G. Yoon, J. Korean Phys. Soc. \textbf{64}, 11 (2014).

\end{thebibliography}
\end{document}